\documentclass[twoside,9pt]{article}
\usepackage{extsizes}
\usepackage[super,sort&compress,comma]{natbib} 
\usepackage[version=3]{mhchem}
\usepackage[left=1.5cm, right=1.5cm, top=1.785cm, bottom=2.0cm]{geometry}
\usepackage{balance}
\usepackage{mathptmx}
\usepackage{sectsty}
\usepackage{graphicx} 
\usepackage{lastpage}
\usepackage[format=plain,justification=justified,singlelinecheck=false,font={stretch=1.125,small,sf},labelfont=bf,labelsep=space]{caption}
\usepackage{float}
\usepackage{fancyhdr}
\usepackage{fnpos}
\usepackage[english]{babel}
\addto{\captionsenglish}{%
  
}
\usepackage{array}
\usepackage{droidsans}
\usepackage{charter}
\usepackage[T1]{fontenc}
\usepackage[usenames,dvipsnames]{xcolor}
\usepackage{setspace}
\usepackage[compact]{titlesec}
\usepackage{hyperref}

\usepackage{epstopdf}

\begin{document}

\title{\textbf{Programming stiff inflatable shells from planar patterned fabrics}} 

\author{Emmanuel Si\'{e}fert, Etienne Reyssat, Jos\'{e} Bico,and Beno\^{i}t Roman  \\
Laboratoire de Physique et M\'{e}canique des Milieux H\'{e}t\'{e}rog\`{e}nes \\
CNRS, ESPCI Paris, Universit\'e PSL, Sorbonne Universit\'e, Universit\'e de Paris,
F-75005, Paris, France}

\maketitle

 \abstract{Lack of stiffness often limits thin shape-shifting structures to small scales. 
The large in-plane transformations required to distort the metrics are indeed commonly achieved by using soft hydrogels or elastomers. We introduce here a versatile single-step method to shape-program stiff inflated structures, opening the door for numerous large scale applications, ranging from space deployable structures to emergency shelters. This technique relies on channel patterns obtained by heat-sealing superimposed flat quasi-inextensible fabric sheets. Inflating channels induces an anisotropic in-plane contraction and thus a possible change of Gaussian curvature. Seam lines, which act as a director field for the in-plane deformation, encode the shape of the deployed structure. We present three patterning methods to quantitatively and analytically program shells with non-Euclidean metrics. In addition to shapes, we describe with scaling laws the  mechanical properties of the inflated structures. Large deployed structures can resist their weight, substantially broadening the palette of applications. 
}

\section{Introduction}
As Carl Gauss demonstrated, curving  a planar surface  in two simultaneous directions requires changing the metrics, i.e. the distances between material points along the surface. 
Nature displays numerous examples of such shape changes induced by  non-uniform growth, which for instance dictates the shape of plant leaves \cite{Nath03, Rolland03} or of our organs \cite{Savin11}. Bio-inspired shape shifting structures have been developed for applications in
tissue engineering \cite{Gao17}, biomedicine \cite{cianchetti2018biomedical}
or drug delivery \cite{sitti2018miniature}. 
In order to achieve the metric distortion required for complex shape morphing, several actuation strategies have been intensively investigated in the last decade, ranging from
swelling hydrogels~\cite{Klein07,Kim12,erb2013self, Gladman16}, liquid crystal elastomers~\cite{warner19,Aharoni18} to, more recently,  dielectric~\cite{bense17,clarke19}  or
pneumatic~\cite{Siefert18} elastomers.
These different solutions rely on basic scalar stimuli, respectively temperature, UV-light, electric field and pressure.
However, their fabrication involves relatively complex processes: control of reticulation rate in hydrogels, precise control of the  orientation of the nematic director field in liquid crystal elastomers, multi-layered electrodes for dielectric elastomers and precise 3D-printed molds for {\it baromorphs}. 
These objects are moreover inherently soft (Young modulus  typically under 1 MPa), hindering applications to human size objects, architecture or space industry.
 \begin{figure*}[!ht]
  \centering
    \includegraphics[width=\textwidth]{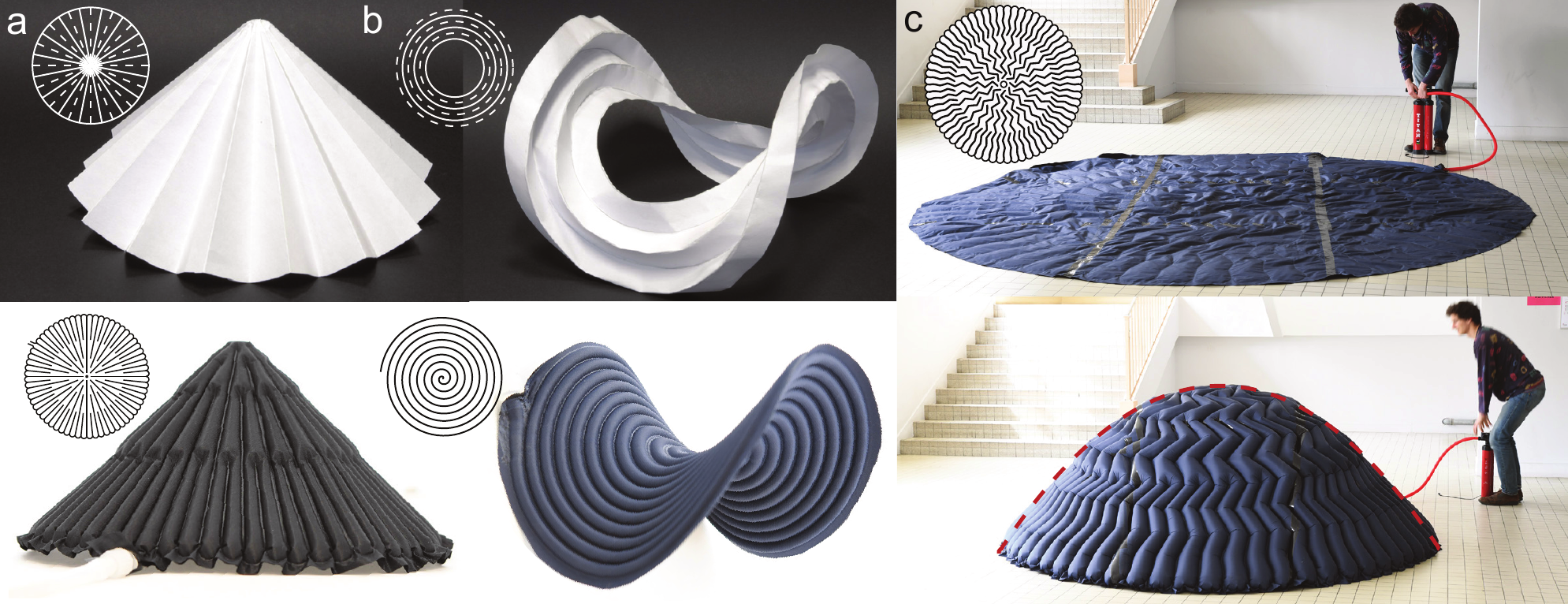}
  \caption{ Origami-inspired design of Gaussian morphing fabrics structures. 
  (a) Paper origami cone made with alternate mountain (solid lines) and valley (dashed lines) radial folds.  Inflated analogue composed of radial seams.
 (b)  Concentric circular folds induce the formation of a saddle shape. The same  anti-cone with seams along an Archimedean spiral.
 (c)  A $3\,$m-wide and $1.2\,$m-high paraboloid structure with a Miura-ori type of pattern, fitting closely to the target shape (dashed line), sustains  its own weight without any significant deflection. 
 } 
       \label{fig1}
\end{figure*}
 \begin{figure}
  \centering
    \includegraphics[width=0.5\textwidth]{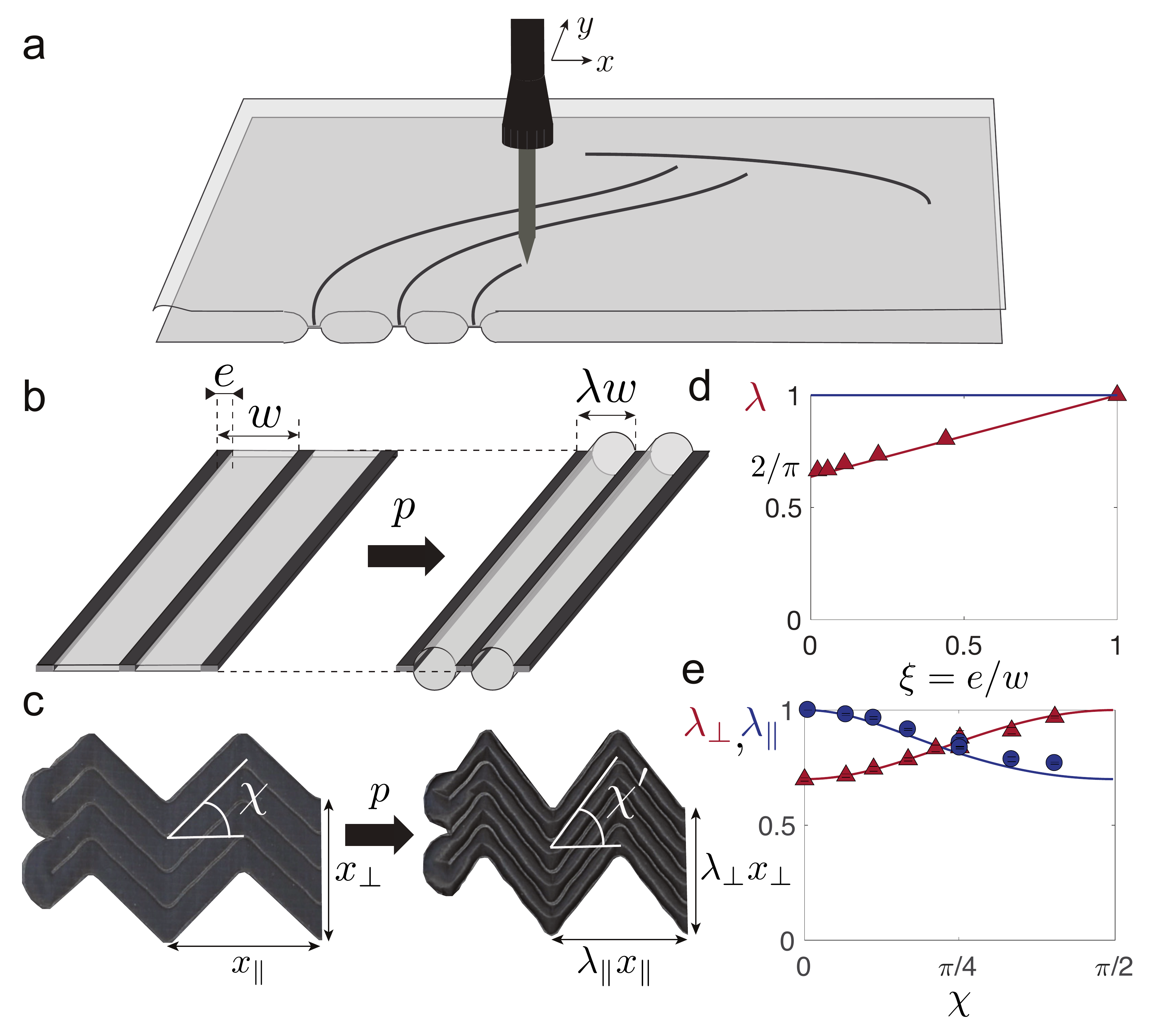}
  \caption{ In-plane metric distortion. 
(a), Two flat superimposed fabric sheets are heat-sealed along any desired path with a heating head mounted on an XY-plotter.
 (b), Upon inflation, the cross section between two locally parallel seams distant by $w$ becomes circular, causing an in-plane contraction of magnitude $\lambda= 2/\pi$.
(c), Varying the relative width $\xi = e/w$ of the seam line, homogenised contraction ratios ranging from $2/\pi$ to 1 can be obtained (red triangle : experiments; continuous line : model)
(d), Deformation of inflated ``zigzag" patterns inspired by miura-ori tessellation of incident angle $\chi$. The orientation $\chi$ of the zigzags increases to a value $\chi'$ upon inflation.
(e), Principal contraction ratios parallel (blue circles) and perpendicular (red triangles) to the average channel direction measured experimentally as a function of the zigzag angle $\chi$. Solid lines correspond to the model (Eq. \ref{eq:lambdapara} \& \ref{eq:lambdaperp}) with $\lambda=0.7$.
 } 
       \label{fig2}
\end{figure}

\begin{figure*}[!ht]
  \centering
    \includegraphics[width=1\textwidth]{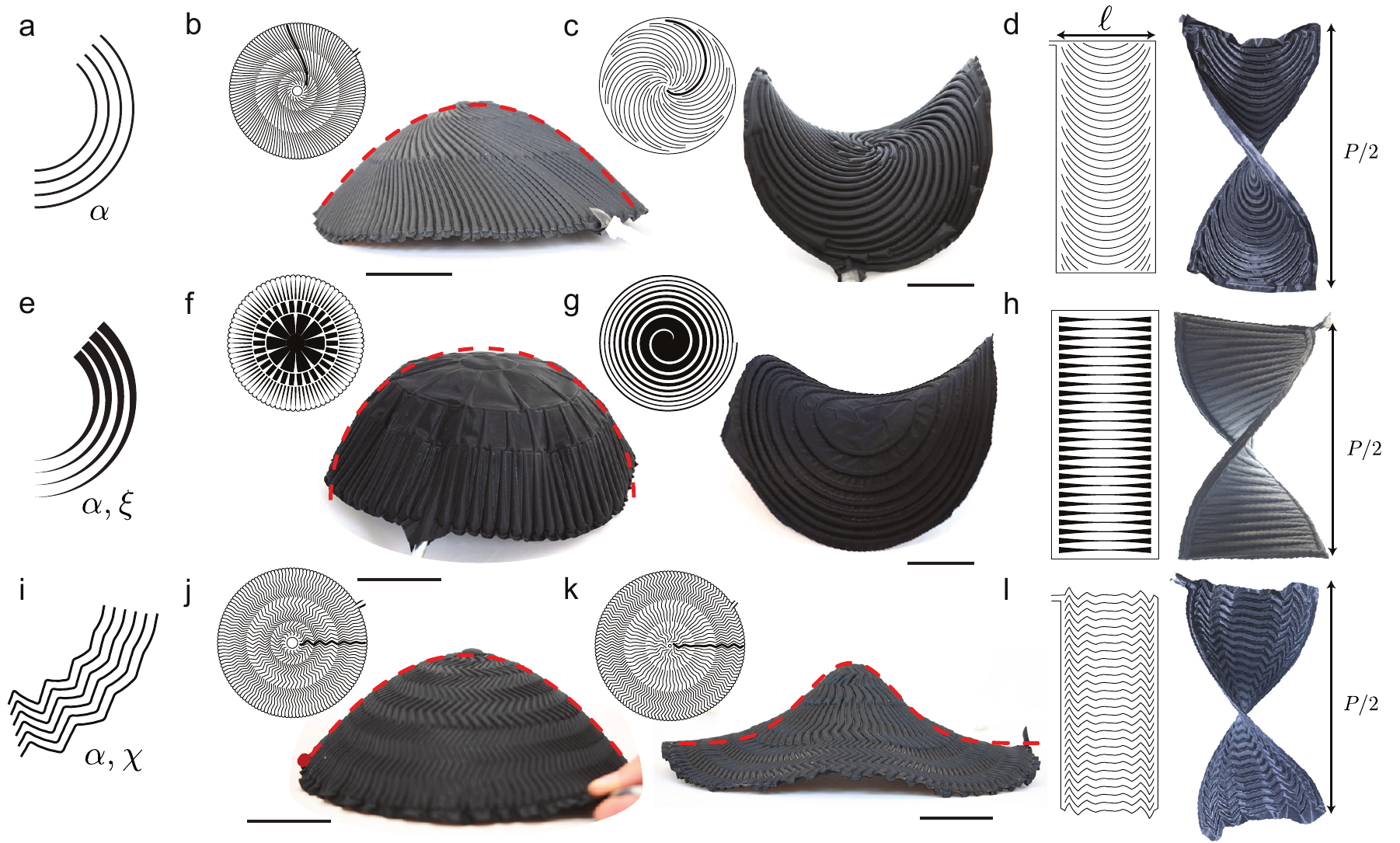}
      \caption{ Three metric distortion strategies.  Simple geometric surfaces are programmed with corresponding seam patterns in insets. 
      (a), Curved seam lines changing the orientation $\alpha$ of the in-plane contraction:
      (b), paraboloid, (c), saddle of constant negative Gaussian Curvature and (d), helicoid.
      (e), Variation of the contraction rate $\lambda$ through the variation of the relative seam width $\xi$, in addition to the control of the orientation $\alpha$ of the seam:
      (f), hemisphere; (g), saddle; (h),  helicoid.
      (i), Zigzag patterns with both the orientation $\alpha$ and the angle $\chi$ of the zigzags as degrees of freedom to distort the metric:
      (j), paraboloid, (k), Gaussian shape, (l), helicoid.
      For axisymmetric shapes, the red dashed lines correspond to the programmed target profiles.
      Each helicoid is programmed to make half a turn (pitch $P$ twice longer than the inflated structure). Scale bars: $5\,$cm.
      }
       \label{fig3}
\end{figure*}
Other strategies to shape plates through a modification of the apparent metric  without  significantly stretching the material rely on cuts (kirigami)~\cite{dias17,castle14,sussman2015algorithmic}, folds  (origami)~\cite{dudte16,zhao2018approximating}, or internal hinges~\cite{Guseinov20}, 
which allow for stiff materials to be used (e.g. polymers or even metals).
Recently, pneumatic pressurization or vacuum have been introduced as a mean  to actuate origami structures~\cite{martinez12,overvelde16,li17,kim19}. However, manufacturing still involves the complex folding of the structure prior to actuation. 
Despite recent advances, the actuation of folds remains indeed complex, prone to errors \cite{Na15,miskin2018graphene} and intrinsically soft.

Here, we present an alternative concept to transform initially flat sheets into stiff and lightweight inflatable shells (Fig.~\ref{fig1}) using a versatile and scalable manufacturing technique \cite{ou16}:
two flat superimposed sheets (typically made of thermoplastic coated fabric) are heat-sealed together along seam lines. 
In contrast with mylar balloon structures where thin sheets are sealed along their edges~\cite{Siefert19pnas}, locally parallel seam lines define here a network of channels over the whole area of the sheets (Fig.~\ref{fig2}a, see ESI$\dag$ and Supplementary movie 1 for details on the manufacturing).
In-plane contractions distort metric in a non-Euclidean way, leading to the buckling of the structure into 3D shapes.
This transformation is programmed by the specific pattern of the network. 
Inflation induces local bending of the sheets (Fig. \ref{fig2}b-c) and an apparent in-plane contraction perpendicular to the channels that we use as an average metric distortion.

In elegant {\it aeromorph} structures~\cite{ou16}, pressurized sheets with specific heat-sealed patterns fold into bistable soft hinges, programming essentially the extrinsic bending of the sheet rather than the intrinsic curvature (i.e., the metric).
Our proposed strategy is closer to tessellated origami:
seam lines are equivalent to the positions of valley folds, with mountain folds in between (Fig.~\ref{fig1}a). 
As a simple example, radial folds in a paper disk lead to the formation of a cone.
Similarly, an inflatable structure with radial seams morphs into a conical shape. 
Conversely, both origami with circular folds and its inflatable analogue with nearly azimuthal seams (Fig. \ref{fig1}b and Supplementary movie 2) buckle into anti-cones \cite{dias12}. 
Contrary to standard origami, the deployment of the 3D structure is here spontaneous upon inflation and does not require tedious mechanical actuation of individual folds. 
Moreover, rigidly and flat-foldable origami tessellations involve soft deployment modes, and cannot be stiff, 
whereas the effective folding angle in our inflatable structures, corresponding to the local contraction rate $\lambda$, is fixed by volume maximization in the highly bendable regime (Fig. \ref{fig2}b-e). 
The shape is obtained by the 2D patterning of flat sheets, not through the complex assembly of multiple patches, as in common inflatable structures \cite{Skouras2014}.
Internal pressure $p$ also provides stiffness to the resulting structure as in other large scale inflatables, such as fabric air beams \cite{comer63}, stratospheric balloons \cite{pagitz07} or even playground castles and architectural buildings \cite{pneumatic}.
We present and rationalize three different ways to distort the metrics and propose analytic procedures to program simple geometric shapes. We then discuss the mechanical properties of such Gaussian morphing fabric structures.

\section{Patterning strategies and axisymmetric shape programming}
In our approach, the pattern is locally made of parallel stripes of width $w$, with seam lines of width $e$ (Fig. \ref{fig2}b). 
Upon inflation, the sheet bends perpendicularly to the seam lines to generate a tubular cross section for sufficiently large pressures ($p\gg Et^3/w^3$), where $t$ is the thickness of the sheet and $E$ its Young modulus.
Owing to the quasi-inextensibility of the fabric sheets ($p\ll Et/w$), this change in cross section leads to an effective in-plane contraction perpendicular to the stripes. 
Taking the thickness $e$ of the seam line into account (but neglecting the effect of seam curvature \cite{Siefert19pnas}),
the effective contraction factor reads: 
\begin{equation}
 \lambda (\xi)= \frac{2}{\pi}(1-\xi)+\xi   
\end{equation}
where $\xi=e/w$ is the relative seam thickness, in very good agreement with experimental measurements (Fig. \ref{fig2}c). 
Conversely, no length change is observed along the seam lines. 
This direction may thus be seen as a director field for the anisotropic metric distortion of magnitude $\lambda$ perpendicular  and $1$ parallel to the lines.
Following the framework developed for liquid crystal elastomers \cite{warner19,Aharoni18}, the metrics of the inflated structure can be written as:
\begin{equation}
a(u,v)=R(\alpha(u,v))^t\begin{pmatrix}
1&0\\
0&\lambda^2\\
\end{pmatrix}
R(\alpha(u,v))
\end{equation}
where $(u,v)$ is a parametrization of the plane, $\alpha$ is the local angle of the director field, $R$ the matrix of rotation and $\lambda$ the contraction rate perpendicular to the channels. Interconnectivity between the channels is ensured by small apertures (smaller than the typical width $w$) at the ends of the seam lines, thus having limited influence on the local contraction governed by volume maximization.
Note that in contrast to nematic elastomers, the contraction rate, $\lambda$ can be varied in the range $[2/\pi,1]$ by tuning $\xi$. However, as the structure is symmetric through the thickness, the extrinsic curvature cannot be programmed.

We now illustrate how this metric distortion strategy can lead to a variety of stiff non-Euclidean shapes upon inflation. 
In the radial seam pattern shown in Fig. \ref{fig1}a, inflation induces a perimeter contraction of amplitude $\lambda$, which leads to the buckling of the disk into a cone of angle $2\arcsin{\lambda}$.
In a nearly azimuthal pattern (archimedean spiral), radii are contracted by $\lambda$, and the structure buckles into an {\it anti-cone} \cite{Dervaux08} with an excess angle $2\pi(\lambda^{-1}-1)$ (Fig. \ref{fig1}b and Supplementary movie 2). Both cone and anticone have a flat metric everywhere except at their apex, where  Gaussian curvature is localized. 
General axisymmetric shapes with distributed Gaussian curvature are programmed by varying the angle $\alpha$ of seam lines with respect to the radial direction (Fig. \ref{fig3}a-d) \cite{warner19, siefertwarner20}, while keeping the seam width and thus the contraction $\lambda$ nearly constant throughout the plane. As in liquid crystal elastomers~\cite{warner19}, the angle $\alpha$ fully determines the ratio of azimuthal versus radial contraction along the inflated shell. 
Arbitrary axisymmetric shapes, e.g. a paraboloid (Fig.~\ref{fig3}b), or structures with constant negative Gaussian curvature (Fig. \ref{fig3}c) can be achieved (see \cite{siefertwarner20} for details on the programming procedure). 
The same method is applied in a cartesian coordinate frame \cite{Aharoni14} to program a helicoid of pitch $P=4.5\,\ell$, where $\ell$ is the width of the deflated ribbon (Fig. \ref{fig3}d and Supplementary movie 3). 
More generally, the design of the director field can in principle be used to generate complex surfaces following the analogy with shape-programmed liquid crystal elastomers~\cite{Aharoni18,griniasty19}.

Nevertheless, we may also take advantage of the possible variation of the contraction $\lambda\in [2/\pi,1]$ by adjusting the width of the seams (Fig.~\ref{fig3}e), which offers an additional degree of freedom in the metric distortion and opens additional shape programming strategies. 
A hemisphere can for instance be  programmed  with radial seams of varying width, using equation (1) and following the simple geometric rule $\lambda(r)=\frac{R}{r}\sin{(r/R)}$,
 where $r$ is the radial coordinate in the flat state and $R$ the programmed radius of curvature of the dome (Fig. \ref{fig3}f). 
 The same method is applied with nearly azimuthal seams of decreasing width to program a saddle of constant negative Gaussian curvature (Fig.~\ref{fig3}g and Supplementary movie 2), or a helicoid with straight parallel seams (Fig. \ref{fig3}h and Supplementary movie 3; see ESI~$\dag$ for more details on the programming).

Another strategy to gain a second degree of freedom in the metric distortion
is to design zigzag patterns that are reminiscent of {\it Miura-ori} origami tessellations.
 In addition to the main direction $\alpha$ of the channels, the characteristic angle $\chi$ of the zigzags (Fig. 2d) may be chosen. $\chi$ controls the ratio of the contractions $\lambda_\parallel$ and $\lambda_\perp$, respectively along and  perpendicular to the average channel direction (Supplementary movie 4).
 Inflating the structure induces a geometrical change from $\chi$ to $\chi'= \arctan(\tan{\chi}/\lambda)$.
Using simple geometric considerations, we retrieve the average contraction rates:
\begin{eqnarray}
\lambda_\parallel=\cos{\chi'}/\cos{\chi}=\dfrac{\lambda}{\sqrt{\sin^2{\chi}+\lambda^2\cos^2{\chi}}}\label{eq:lambdapara}\\
\lambda_\perp=\lambda\cos{\chi}/\cos{\chi'} =\sqrt{\sin^2{\chi}+\lambda^2\cos^2{\chi}}
\label{eq:lambdaperp}
\end{eqnarray}
which are in quantitative agreement with experimental measurements (Fig.~2e). Note that changing the angle $\chi$ of the zigzag does not impact the overall area contraction upon inflation, which remains equal to $\lambda=\lambda_\perp\lambda_\parallel$. Indeed, every air channel (from a zig or a zag) locally contracts uniaxially by an amount $\lambda$.
Zigzag patterns can thus be viewed as a global isotropic area contraction followed by an in-plane shear varying both in direction (orientation $\alpha$ of the zigzag) and intensity (angle $\chi$). 
Similarly to the computational techniques used with curved seams, zigzags (Fig. \ref{fig3}i) may be used to quantitatively program axisymmetric shapes (Fig. \ref{fig3}j,k and Supplementary movies 5,6) or a helicoid (Fig.~\ref{fig3}l and Supplementary movie 3) (see ESI$^\dag$  and \cite{Gao20} for more details on the design of the seam networks).
\begin{figure}[!ht]
  \centering
    \includegraphics[width=0.5\textwidth]{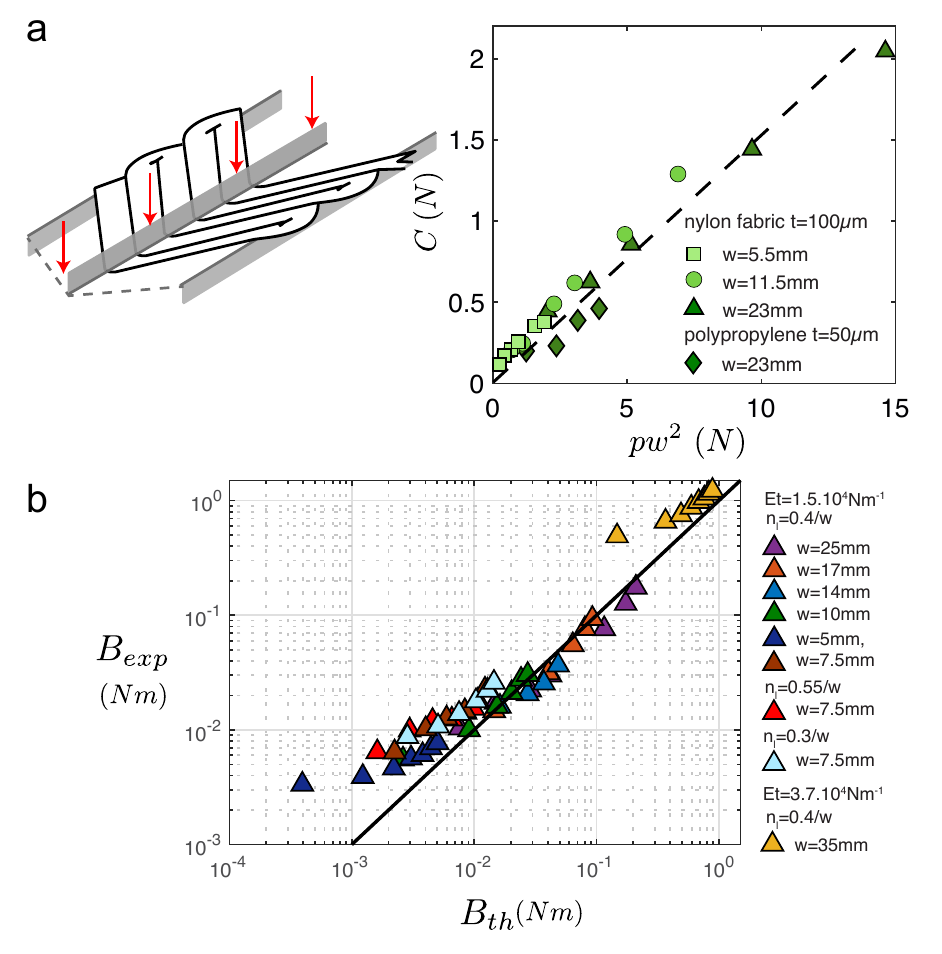}
      \caption{Stiffness.
       (a) Inflated zigzag patterns bend preferentially along pseudo-hinges defined by junctions between zigs and zags. A three-point bending test provides measurements of the hinge stiffness resulting from volume variations. 
       The hinge stiffness increases linearly with pressure. (b) Experimental value of the homogenized bending stiffness versus the prediction of the minimal model given by equation 6. 
      }
       \label{fig4}
\end{figure}

\section{Stiffness of inflatable shells}
Beyond geometry, such deployed structures are shells made of a collection of inflated beams and locally present highly anisotropic stiffness both for bending and stretching.
In the regime of interest, the transverse bending stiffness per unit width  of an array of parallel inflated beams scales as $B_\perp\sim Etw^2$, as in other inflatable structures \cite{comer63}.
 Although $B_\perp$ barely depends on the applied pressure (as long as the air beam adopts the optimal circular section, i.e. $p\gg Et^3/w^3$), the maximum moment per unit width an array of beams can sustain without failing strongly depends on pressure ~\cite{calladine1989theory,seide1961buckling,guo20} and typically scales as $pw^2+Et^2$  (see ESI~$\dag$ for more details).
Conversely, the bending stiffness along seam lines is proportional to $Et^3$ and is therefore  orders of magnitude smaller than in the transverse direction : seams act as soft hinges between rigid inflated tubes. As a consequence, long straight seams favor floppy modes (e.g. the cone and the dome respectively illustrated in Figs.~\ref{fig1} and \ref{fig3}f bend easily along radial lines) whereas curved and narrow seams promote the global stiffness of the inflated structure. This strong stiffness anisotropy has a major impact on the shape selection among isometric embeddings of the target metrics: for instance, a helicoid (Fig. \ref{fig3}h) is selected rather than a catenoid, since it does not require the bending of the horizontal beams.
Zigzag patterns appear as an interesting strategy to ensure the variation of the direction of the air beams at a mesoscale, while inducing stronger and more isotropic mechanical stiffness of the inflated structure. Since the envelope does not extend, the stretching modulus per unit width of pressurized patterns is related to variations of the enclosed volume and scales as $pw$ in both directions  (see ESI$\dag$ for more details). 
 Moreover, sharp changes in the direction of the seam induce, in the inflated state, compressive folds in the vicinity of the junctions between ``zigs'' and ``zags''~\cite{Siefert19pnas}.
 In this region, the zigzaging beams may therefore deform through folding or unfolding the membrane without material strain. Upon bending, the curvature of such zigzag structures localizes at these softer spots. A kinking angle $\phi$ is achieved by the work $p\Delta V$ against the imposed pressure $p$, where  the volume change $\Delta V$ is quadratic in $\phi$. 
The localized linear hinge stiffness per unit width $C= p\,\partial^2(\Delta V)/\partial\phi^2$ is therefore proportional to $p$: 
 \begin{equation}
  C = pw^2 {\cal F}(\chi) 
  \label{eq:C}
 \end{equation}
where $w$ is the only local length scale and $\cal F$ a function of the zigzag angle $\chi$.
Experimental measurements using a three point bending test (Fig. \ref{fig4}a) are consistent with this prediction: $C$ is linear with the pressure and appears independent of the sheet material properties.

Hinges can thus sustain a maximum moment that scales as $pw^2$ (as for straight beams). They also allow for much larger curvatures before an overall collapse, as many kinks are distributed over the structure. 
Hence, inserting hinge-like singularities typically reduces the local bending stiffness but also prevents the formation and localization of a single catastrophic  kink in the structure.
The global bending rigidity of zigzag patterns may be seen as a collection of beam components of stiffness $B_0=Ew^{2}t/(4\pi)\cos^2{\chi}$ and hinges of rigidity $C$. The homogenized rigidity thus reads:
\begin{equation}
    B_\perp=(B_0^{-1}+n_lC^{-1})^{-1}
    \label{eq:stiffscale}
\end{equation}
where $n_l$ is the number of direction changes per unit length. 
In the typical regime of interest, both terms are of the same order of magnitude and none may be neglected. 
Despite the strong simplifications implied in our description, equation~(\ref{eq:stiffscale}) provides a satisfying scaling law (using the prefactor found experimentally in Fig.~\ref{fig4}a for $
C$), especially at large enough pressures (where the assumption of circular cross section of the beams is verified).

All in all, the maximum moment per unit length the structure can sustain scales classically as $pw^2$. The typical maximum size of such inflatable structures $L_{max}$ at which they can sustain their own weight scales thus as $L_{max}\sim [{pw^2}/(\rho gt)]^{1/2}$.
For typical values ($E\sim10^9\,$Pa, $w\sim 1\,$m, $\rho\sim 10^3\,$kg.m$^{-3}$, $g\sim 10\,$m.s$^{-2}$, $t\sim10^{-3}\,$m and $p\sim 10^4\,$Pa), $L_{max}$ amounts to tens of meters. 
Architectural shape-morphing structures are thus reachable with this strategy. In order to highlight the high stiffness to weight ratio of such structures, a $4\,$m wide structure ($3\,$m in the inflated state) has been manufactured using an ultrasonic sewing machine (Fig.~\ref{fig1}c and Supplementary movie 7).

\section{Conclusion}
Gaussian morphing fabrics constitute a versatile and simple technique to produce stiff shape-morphing pneumatic structures with well-defined programmable shapes. 
The manufacturing process is scalable and architectural size structures are within reach. 
Several patterning strategies - lines, seams of varying thickness and zigzags - have been introduced, allowing for one or two degrees of freedom in the metrics prescription. Although a wide variety of shapes can be programmed analytically, the general inverse problem, i.e. programming a pattern of seam lines such that the inflated structure deploys into a desired target shape, has to be solved numerically \cite{Aharoni18,konakovic16,boley19,griniasty19} and is beyond the scope of this article, since state-of-the-art techniques \cite{Aharoni18} fail due to the strong mechanical anisotropy of the structures.

Numerous extensions are possible for practical applications: several layers may be stacked with specific welding patterns between two consecutive fabric sheets, each pattern coding for one specific shape (Supplementary movie 8). As the fabric used remains unstretched upon deployment, the structures may support  metallic tracks for shape changing electronic devices~\cite{rogers2010materials}. Other actuation strategies may also be envisioned, such as hydrogel swelling inside the structure (Supplementary movie 9).
In order to bias the symmetry and promote a preferred deployment direction, fabric sheets of different thicknesses may finally be used. Altogether, our study offers a simple manufacturing platform where stiff shape-morphing structures are anticipated to find new innovative applications at human and architectural scale, ranging from rehabilitation medical tools to emergency shelters.

\section*{Conflicts of interest}
There are no conflicts to declare.

\section*{Acknowledgements}
We thank Tian Gao, Maïka Saint-Jean and Manon L'Estim\'{e} for technical help and Mark Pauly, Mark Warner, Julian Panetta and Antonio DeSimone for fruitful discussions. This work was supported by ANR SMArT.



\balance


\bibliography{article} 
\bibliographystyle{unsrt} 
\end{document}


\maketitle

\newpage

\tableofcontents

\newpage
\section{Movies}

{\bf Supplementary Movie 1 \\
Manufacturing process of orifusen}\\
Two heat-sealable sheets are superimposed and fixed in the working area of a CNC-machine. A heating head (soldering iron) is mounted on the machine and directly plots the desired welding pattern (video accelerated x80). The boundaries are then cut with scissors and a connector is inserted at the inlet 
of the structure. The structure is connected to a pressure supply (e.g. an inflating bulb) and buckles upon inflation into a 3D shape prescribed by the welding pattern.\\\\
{\bf Supplementary Movie 2 \\
Anti-cone and saddle-shaped orifusen.}\\
An archimedean spiral pattern (and thus nearly azimuthal welding direction) induces the radial contraction of the structure upon inflation, which buckles into an anti-cone. The structure has a diameter of 15 cm and is made of two superimposed Nylon fabric sheets (70den, one-sided TPU-coated, 170 g/sqm) from Extremtextil.   
Varying the relative welding width $\xi$ in a nearly azimuthal channel pattern controls the local radial contraction rate $\lambda$, and a saddle of constant negative Gaussian Curvature may be programmed. The structure has a diameter of $25\,$cm and is made of two superimposed Ripstop-Nylon fabric sheets ($40\,$den, one-sided TPU-coated, $70\,\mathrm{g/m}^2$) from Extremtextil.
\\\\
{\bf Supplementary Movie 3 \\
Three different strategies to program a helicoid.}\\
Three strategies may be used to distort the planar metric: changing only the orientation of the channels, playing additionally with the relative width of the seam lines, and making zigzag patterns of varying angle. Three helicoids are programmed using each method. The structures have a width of approximately $15\,$cm and are made of two superimposed Ripstop-Nylon fabric sheets ($40\,$den, one-sided TPU-coated, $70\,\mathrm{g/m}^2$) from Extremtextil.  \\\\
{\bf Supplementary Movie 4 \\
 Influence of the zigzag angle $\chi$ on the principal contractions. } \\
 The variation of the angle $\chi$ of the zigzags induces a change in the ratio between the contraction along ($\lambda_\parallel$) and perpendicular  ($\lambda_\perp$) to the averaged zigzag direction.
 \\\\
{\bf Supplementary Movie 5 \\
Dome programmed with a zigzag pattern.}\\
A dome of constant positive curvature can be programmed by varying the angle of zigzags (whose average direction is radial).  The structure has a diameter of approx. $25\,$cm and is made of two superimposed Ripstop-Nylon fabric sheets ($40\,$den, one-sided TPU-coated, $70\,\mathrm{g/m}^2$) from Extremtextil. \\\\
{\bf Supplementary Movie 6 \\
Gaussian shape programmed with a zigzag pattern.}\\
The structure has a diameter of approx. $25\,$cm and is made of two superimposed Ripstop-Nylon fabric sheets ($40\,$den, one-sided TPU-coated, $70\,\mathrm{g/m}^2$) from Extremtextil. \\\\
{\bf Supplementary Movie 7 \\
Inflation of a 4~m wide orifusen structure, programmed to shape into a paraboloid.}\\
The structure is made of  two superimposed Nylon fabric sheets ($70\,$den, one-sided TPU-coated, $170\,\mathrm{g/m}^2$) from Extremtextil. The welding pattern has been made using an ultrasonic sewing machine. \\\\
{\bf Supplementary Movie 8 \\
Three layers orifusen.}\\
Stacking three layers of fabric with specific welding patterns between the top-middle and middle-bottom layers enables to get two different shapes from one structure (when the top-middle pattern is inflated and when the middle-bottom one is inflated). Here, we reproduce the different shapes of the cap of algae {\it Acetabularia Acetabulum} during its growth \cite{Dervaux08}. \\\\
{\bf Supplementary Movie 9 \\
Hydrogel actuated orifusen}\\
Two layers of standard cotton  fabric can be simply sewed together along a specific pattern and small hydrogel beads (``water beads'') can be inserted in the structure, which remains mostly flat. When immersed in water, the beads swell and deploy the structure into its target shape. The structure has a diameter of approx. $18\,$cm. Total duration of the movie is one hour.  \\\\
\section{Structure fabrication}
 
The fabrication of Gaussian morphing fabrics is relatively simple, since the manufacturing process is completely 2D.
\paragraph*{Material}
Any thermoplastic material may be potentially used, including among others Mylar, polypropylene, polyethylene. Nevertheless, nylon textiles coated with thermoplastic urethane (TPU) matrix are best suited. They present the advantage to be easily and strongly sealed at relatively low temperature (around $150\,^\circ$C). At such temperatures, only the TPU matrix melts and not the nylon textile, thus avoiding undesired puncture of the sheet while welding. Such fabrics are also stiff in  the direction of nylon fibers (average Young's modulus of typically $1\,$GPa) while allowing some shear. This slight shear prevents the apparition of kinks and localized folds in the structure.

The structure is primarily made of two superimposed thermoplastic sheets that should be sealed together along a specific pattern. Various simple techniques may be used to weld the two sheets. 
\paragraph*{Heat printing} A soldering iron may be mounted on an XY-plotter or a CNC-machine to directly "print" the heat-sealed seams pattern on the two superimposed sheets (Fig. 2a in the main document).
We used the XY-plotter from Makeblock and a CNC Workbee from Openbuilds.
The temperature, pressure and displacement velocity of the heating head are tuned in order to obtain a strong bonding between the two layers. These technical settings are however sensitive to the thickness and material properties of the sheets. 
The rationale behind the choice of parameters is the following: the temperature should be high enough to melt the TPU, the displacement velocity should be slow enough to ensure heat diffusion through the thickness of the sheet and the load applied on the tip should be high enough to impose a strong bonding (but too much loading may damage the sheets). In order to control the load at the tip, a slide is mounted on the Z axis of the plotter, the soldering iron being attached to the slide.
Additional weights 
may be implemented to the slide to adjust the load. 
For the TPU-impregnated fabrics used in the illustrated examples, 
we set a typical speed of $150\,$mm/min, a total weight of approximately $500\,$g and a head temperature at $220\,^\circ$C. The diameter of the head is around $1\,$mm.

If the sheet is coated on both sides with thermoplastic material, a sheet of baking paper is placed on top of the structure during the welding in order to protect the structure and to avoid the adhesion of the melted material on the printing head.

\paragraph*{Heat press} Alternatively, the pattern of air channels may be laser cut in baking paper so that the designed seam lines are cut out from the baking paper sheet.
The resulting pattern is then placed between both thermoplastic fabric sheets; the whole sandwich is heat-pressed between two additional baking paper sheets to prevent the adhesion of melted thermoplastic on the heat press.
The two fabric sheets are sealed together only where the baking paper has been cut out.
The patterned baking paper remains trapped in the structure but does not play any mechanical role upon inflation.

\paragraph*{Heat press mould}
This method is best suited for serial production of the same structure. A metallic plate is engraved  except at the target location of the seam lines. It serves as a waffle iron in the heat press, welding only the appropriate lines on the surface. This fabrication technique could be directly used for industrial applications, enabling a large and cost-efficient throughput.

\paragraph*{Other methods}
Alternatively, regular sewing is also possible to manufacture the structures, if a sealant film is applied on the seam lines to ensure air tightness. Any impermeable fabric may thus be used with such a fabrication strategy. 
Pressure within the structure may be caused by other means, including hydrogel beads swelling in water. In such a case, the fabric sheets have to be permeable to allow water diffusion in the structure, promoting the shape change (Supplementary movie 9).

\section{Metric distortion strategies}

\subsection{Width of the seam}
In the first presented version of {\it orifusen} structures, inflated shapes only result from the local direction of the seam lines.
An additional degree of freedom in the distortion of the flat metrics  widens the possibilities of shape programming.
A simple solution is to play with the relative seam width $\xi$ to tune the homogenized in-plane contraction perpendicular to the seam direction (see Eq.~1 in the main manuscript). 

\paragraph*{Hemisphere.} 
Consider purely radial airways, that induce azimuthal contraction. 
We define as $u$ the curvilinear coordinate in the inflated structure and $r$ the radial coordinate of the initially flat state.
In the case of radial seams, there is no contraction in the radial direction ($u=r$).
In order to program a portion of a sphere of radius $R$, the evolution of the azimuthal contraction $\lambda_\theta$ should follow:
\be
\lambda_\theta(r)=\frac{R}{r}\sin{\frac{r}{R}}=1-\frac{r^2}{6R^2}+o((r/R)^2)
\ee
The relative seam width $\xi(r)$, may be easily computed by combining this relation with Eq.~1 from the main manuscript. The maximum achievable contraction is $2/\pi$,  which enables us to program a hemisphere out of a flat disk (the equator indeed verifies $r/R=\pi/2$).
The structure does buckle, when inflated, into a bowl shape that is close to the targeted hemisphere (Fig.~3f in the main manuscript). 
Nevertheless, the structure remains very floppy, 
as it may be easily folded along any radial seam.

\paragraph*{Saddle.} In order to program a saddle of constant negative Gaussian curvature $K$, the perimeters should bear excess length: $P(u)=2\pi u(1-Ku^2/6+o(u^2))$.
In the ideal case of purely azimuthal seams, 
perimeters remain unchanged upon activation, but the radii contract. 
At second order in $r$, the contraction should thus reads:
\be
\lambda_r(r)=1+\frac{Kr^2}{6}
\ee
Although the weld lines are not perfectly azimuthal (spiral pattern), the obtained shape is satisfactory (Fig.~3g in the main manuscript) and appears to be stiff because of the curved seams.

\paragraph*{Helicoid.}
Consider parallel seams of varying width. The contraction should be maximal at the center line of the ribbon and can be chosen as $\pi/2$ with no contraction along the edges.
Thus, the maximum admissible ratio $l/p$, where $l$ is the width of the ribbon and $p$ the pitch of the helicoid, reads:
\begin{equation}
    \frac{l}{p}=\frac{1}{2}\sqrt{\frac{\lambda_{max}^{-2}-1}{4\pi^2}}
\end{equation}
where $\lambda_{max}$ corresponds to the maximum contraction, \textit{i.e.}  the minimal value of $\lambda$
($\lambda_{max}$ is bounded by $2/\pi$ in the limit of vanishing thickness of the seam line). 
The variation of the targeted contraction rate $\lambda_\perp$ as a function of the coordinate along the seam $x_\parallel$ (the origin being chosen at the center of the ribbon) can be expressed as:
\begin{equation}
    \lambda_\perp(x_\parallel)=\lambda_{max}\sqrt{1+4\pi x_\parallel^2/p^2}
\end{equation}
Using equation (1) from the main article, the relative width of the seam $\xi$ reads:
\begin{equation}
    \xi(x_\parallel)=(\sqrt{1+4\pi x_\parallel^2/p^2}-1)/(\lambda_{max}^{-1}-1)
\end{equation}
Additionally, two parallel stripes are sealed along the outer edges of the ribbon to ensure bending stiffness along the helicoid direction (Fig.~3h in the main manuscript).

The dome obtained with this strategy is very floppy, whereas the helicoid and the saddle are relatively stiff. 
 Both structures indeed present a boundary that is not contracted along its tangent: the contraction is perpendicular to the boundary, which is stiffened by the presence of the curved air beam. This is possible only for negative curvatures, since the plate may only contract (and not extend) within a fixed boundary length. 
Another condition to obtain a stiff structure is to ensure that the welded portions are under tension. 
As they are very slender, they would indeed buckle for minute compressive forces.

\subsection{Programming axisymmetric shapes with zigzag patterns}
\label{sect:zigzag}

Using zigzag patterns (see main manuscript), simple geometric surfaces with Gaussian curvature may be programmed. 
Consider a target axisymmetric shape obtained by the revolution of the curve  $z=f(\rho)$, where $(\rho, z)$ correspond to the cylindrical coordinates. 
We consider zigzag patterns in the radial direction.
In this case, the remaining degree of freedom is the internal zig-zag angle $\chi(r)$, where $r$ is the radial coordinate in the initially flat disk.
 Starting at the center of the disk ($r=0$), the inflated structure should be locally flat to avoid a singular point, which imposes $\lambda_\perp=\lambda_\parallel$ and thus $\chi(r=0)=1/\sqrt{1+\lambda}$. 
 Making an infinitesimal step $\d r$ on the flat disk thus results in an infinitesimal step: 
 \be
\d u= \lambda_\parallel(r) \d r
\ee
 on the inflated curved surface. At this location, the needed azimuthal contraction  simply reads: 
\be
\lambda_\perp=\rho/r
\ee
 from which we retrieve the appropriate local angle $\chi(r)$.  The infinitesimal step for $\rho$ is set by the geometric relation:
\be
\d u=\sqrt{\d \rho^2 +\d z^2}=\sqrt{1+f'(\rho)^2}~\d \rho
\ee

\noindent
Putting the last three equations together and using the constant area contraction $\lambda_\perp\lambda_\parallel =\lambda$, we retrieve the following expression:
\be
\lambda r \d r=\sqrt{1+f'(\rho)^2}~\rho\d \rho
\label{eq:numerical}
\ee

\noindent
The local angle $\chi(r)$ of radial zigzags follows the same evolution as the seam line orientation angle $\alpha(r)$ in spiral patterns~\cite{siefertwarner20}.

\noindent
Depending on the configurations, Eq.~\ref{eq:numerical} may be solved analytically  or numerically. A paraboloid is shown in Fig.~3j and Fig.~4 from the main manuscript. The shape of the inflated shell quantitatively matches the target profile (in red dashed lines). A Gaussian profile is presented in Fig.~3k from the main manuscript. This target shape is in practice not reachable with the strategy implying the variation of the width of the welded line. The curvature of the profile indeed induces compressive azimuthal forces, which would result in the buckling of the welded portions. 

\section{Structural Stiffness}\label{sec:stiffness}
In this section, we describe the stretching and bending stiffness and maximum admissible moment of straight inflated textile beams and arrays of zigzag beams, in order to  better understand the mechanics of the structures we build.
\subsection{Stiffness of straight lines}

Consider  a pair of facing strips of thickness $t$ sealed along their edges of length $l$ and width $w$, with $t\ll w \ll l$. 
The width of an assembly of $N$ parallel inflated strips is thus $L_0=2Nw/\pi$ .
We define as $w^*= w-e$, the width of the unsealed portion of the fabric (in practice $e\ll w$).
Upon inflation, the structure deforms into an air beam of circular cross section with radius $r=w/\pi$ if the pressure is large when compared with the bending resistance of the strips, i.e. if $p\gg Et^3/w^{*3}$, where $E$ is the Young modulus of the fabric, 
which is the case in our experiments. 
We also assume that the stretching strain induced by pressure is small, i.e. $p\ll Et/w$, 
which is also verified experimentally.
 \begin{figure}[!ht]
  \centering
    \includegraphics[width=1\textwidth]{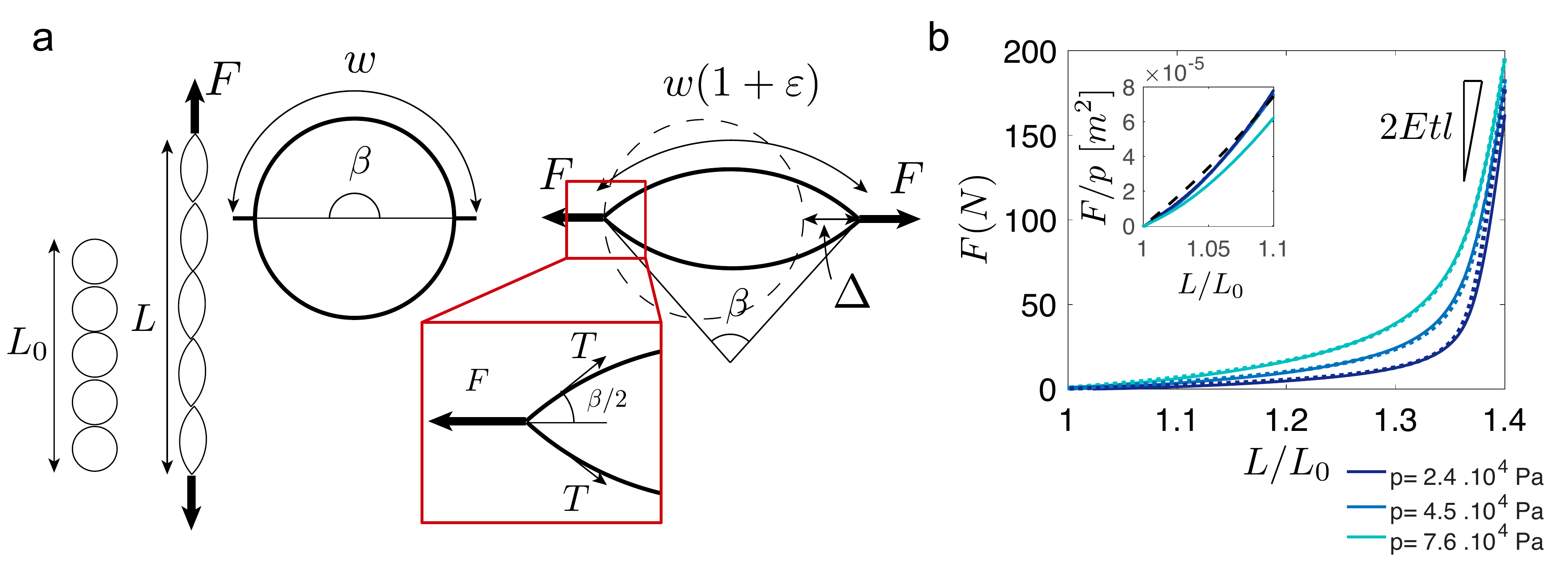}
      \caption{Stretching of the structure perpendicular to the direction of the seam. 
      (a) Schematic cross section of an inflated beam at rest and under traction, with the spanned angle $\beta$, the effective strain at the scale of the sheet $\varepsilon$, the net displacement $\Delta$ and a zoom on the force balance at the seam. (b) Force-displacement curve for various pressures, with experimental measurements in solid lines, and theoretical prediction in dashed lines.
      Inset: rescaled force-displacement curve in the quasi-inextensible regime: the force does not depend on the  properties of the sheet. The structure corresponds to 10 tubes with $w=5.5\,$mm, $e=1.7\,$mm, 
      $E=4.5.10^8\,$Pa, and $t=0.1\,$mm.
      } 
       \label{figstraightstretch}
\end{figure}
\paragraph{Stretching.} Along the beams, the stretching modulus of the inflated structure $Y_\parallel$ simply reads $Y_\parallel = \pi Et$ 
(the potential contribution from pressure scales as $pw$ which is order of magnitudes smaller than stretching the sheet). 
Conversely, stretching the structure in the direction perpendicular to the seams means elongating the cross-section towards a solution that does not maximize the volume. 
Laplace law imposes 
the cross sections of the beams to remain portions of circles of spanned angle $\beta$ and corresponding radius of curvature $R=w^*(1+\varepsilon)/\beta$ (Fig. \ref{figstraightstretch}a), where $\varepsilon$ is the (uniform) strain applied to the structure.
The tension $T$ in the sheet reads:
\begin{equation}
    T=Et\varepsilon=pw^*(1+\varepsilon)/\beta
    \label{eq:tens}
\end{equation}
A simple force balance (Fig. \ref{figstraightstretch}a) gives the following expression for the force $F$ (per unit length):
\begin{equation}
    F=2T\cos(\beta/2)
\end{equation}
The corresponding total displacement $\Delta$ reads:
\begin{equation}
    \Delta=L-L_0=2w^*\left((1+\varepsilon)\,\frac{\sin(\beta/2)}{\beta}-\frac{1}{\pi}\right)
    \label{eq:delta}
\end{equation}
Combining equations (\ref{eq:tens}-\ref{eq:delta}), one may eliminate $T$ and $\varepsilon$, and express both the force and the displacement as a function of $\beta$
\bea
     & F= 2pw^*\,\dfrac{cos(\beta/2)}{\beta-\Sigma}  \\
     & \Delta=2w^*\left(\dfrac{\sin(\beta/2)}{\beta-\Sigma} -\dfrac{1}{\pi-\Sigma}\right)
\eea
where $\Sigma=pw^*/(Et)$ is a small dimensionless number as discussed above. 
These expressions are plotted in Fig. \ref{figstraightstretch}b (dashed lines) and match precisely the force-displacement curves measured experimentally (solid lines) for various pressures. The seam-line width $e$, difficult to measure, has been used as a fitting parameter and yields reasonable values ($1.7\,$mm), close to the estimated measurements.

\noindent
Linearization of the ratio $FL_0/(l\Delta )$ around the unloaded inflated state ($\beta=\pi$) provides an analytical expression for the stretching modulus perpendicular to the seam direction $Y_\perp$ for infinitesimal deformation.
\begin{equation}
    Y_\perp=\frac{2}{\pi}\,pw^*
\end{equation}
For small deformations, the effective stretching modulus does not depend on the fabric and is only set by the pressure and the geometry of the air beams. Dividing the force by the pressure results in a collapse of the force vs. strain curves (Fig. \ref{figstraightstretch}b inset). 
Conversely, the stiffness of the structures reaches the  stiffness $2Et$ of the pair of sheets for large deformations, as illustrated in Fig. \ref{figstraightstretch}b. 

\paragraph{Bending.}For small bending deformations, 
pressure does not play any role, since bending conserves volume at the first order bending strain \cite{comer63}. The bending stiffness $B_\perp$ per unit width of the beam thus corresponds to the bending stiffness of the envelope in the inflated geometry, and reads (we here consider $w^* \simeq w$): 
\begin{equation}
    B_\perp=Ew^{2}t/(4\pi)
    \label{eq:bendstiff}
\end{equation}

 \begin{figure}[!ht]
  \centering
    \includegraphics[width=1\textwidth]{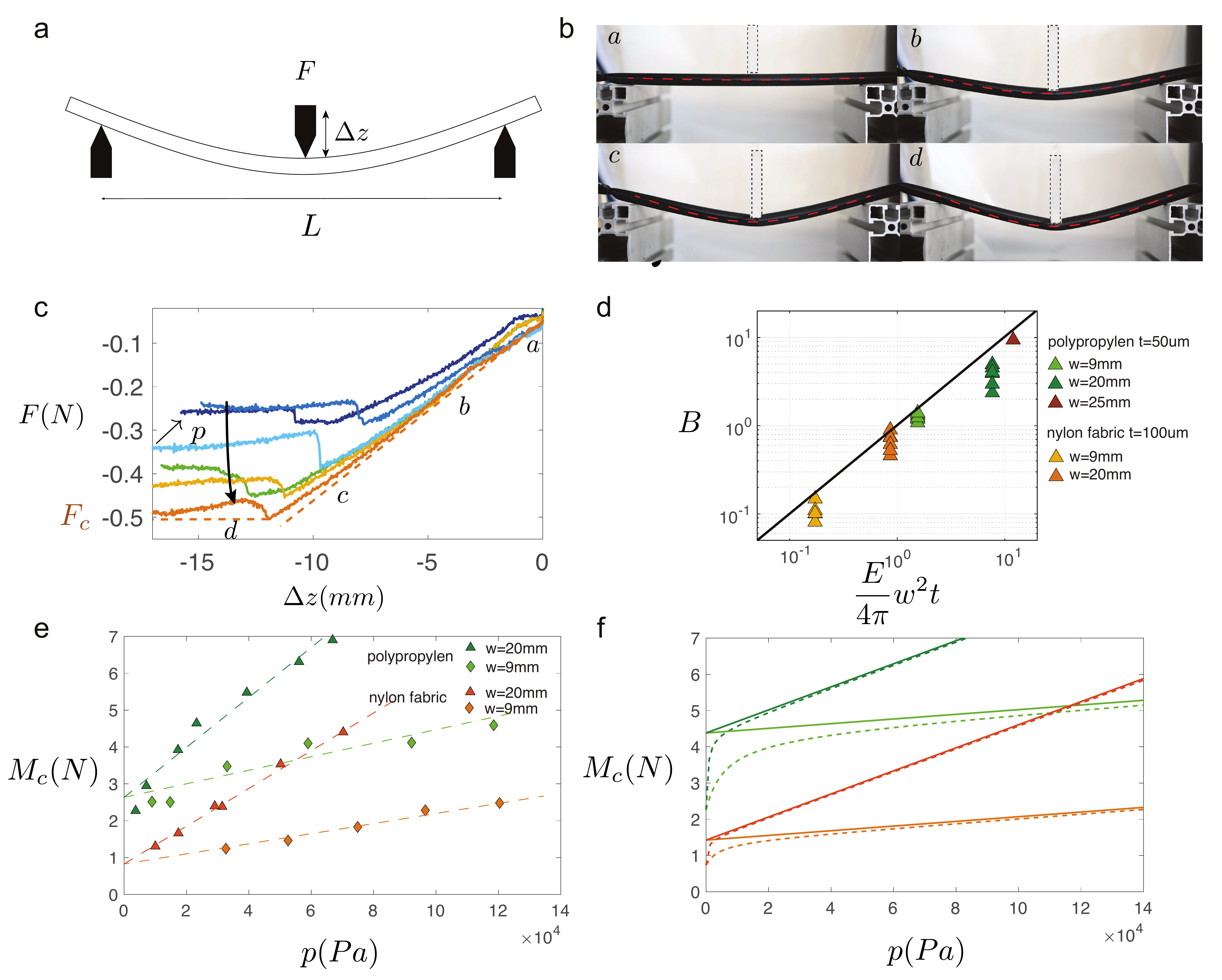}
      \caption{Three-point bending test of a straight inflated beam. (a) Experimental setup. (b)~Pictures for various imposed deflections with the superimposed theoretical profile in red dashed lines. 
      The four pictures, labeled by {\it a, b, c, d}, correspond to different stages of a bending test, the force-displacement curves being shown in (c) for an air beam of deflated width $w=9\,$mm made of a polypropylene sheet of thickness $t=50\,\mu$m and Young's modulus $E=2.2\,$GPa. (d)~Experimental bending stiffness per unit width $B$ as a function of the theoretical bending stiffness of air beams. Experimental (e) and theoretical (f) critical moments per unit width $M_c$ as a function of pressure accounting for ovalization (dashed lines) or not (solid lines).} 
       \label{figbendstraight}
\end{figure}

\noindent
In order to probe this prediction experimentally, slender air beams are 
subjected to a three-points bending test for various pressures, as shown in Fig.~\ref{figbendstraight}(a) and (b). The applied force first displays a linear dependence with the imposed displacement $\Delta z$ (Fig. \ref{figbendstraight}c),
as predicted by classical Euler-Bernoulli beam theory:
\be
\Delta z= \frac{FL^3}{48EI}
\ee

We obtain a fair agreement between the inferred stiffness and its theoretical value $Ew^2t/4\pi$ (Fig.~\ref{figbendstraight}d).\\

Although this linear response is mostly independent of pressure, such an air beam dramatically collapses upon a critical load $F_c$ that strongly depends on pressure, as shown in Fig.~\ref{figbendstraight}c.
While the profile of the beam follows the classical Euler-Bernoulli description, a sharp king is observed when the force reaches $F_c$ (Fig.~\ref{figbendstraight}b).   
The critical moment per unit width $M_c=\pi F_cL/8w$ is plotted as a function of the pressure for various air beams in Fig.~\ref{figbendstraight}e made of different materials and with  cross-sections of various 
widths. 
The dependence of $M_c$ on $p$ appears to be affine, where the slope varies with the width of the cross-section, while the critical moment at vanishing pressure mainly depends only on the properies of the sheet (Fig.~\ref{figbendstraight}e).\\

\noindent
Following Calladine \cite{calladine1989theory} and Seide and Weingarten \cite{seide1961buckling}, the critical torque $M_c$ for which an air beam collapses is attained when the extreme fibre reaches the critical buckling stress $\sigma_c$ in the case of uniaxial compression of a cylindrical shell. The critical stress reads $\sigma_c=\pi E t/(\sqrt{3}w)$. In order to reach this critical compressive stress, the imposed moment must additionally overcome the pressure-induced longitudinal tension in the beam, which classically reads $\sigma_p=pw/2\pi t$.
Hence the maximum bending moment per unit width that the air beam may sustain reads:

\begin{eqnarray}
    M_{c}&=\pi w\frac{t}{2}[\sigma_c+\sigma_p]\\
   M_{c}&=\dfrac{\pi^2 Et^2}{2\sqrt{3}}+\dfrac{pw^2}{4\pi}
    \label{eq:mommax}
\end{eqnarray}

\noindent
The ovalization of the cross section upon curvature, known as Brazier effect, may be disregarded in these examples, since the  applied pressure limits changes of the shape of the cross-section.  The theoretical critical momentum with (dashed lines) and without (solid lines) the ovalization of the cross section as a function of the pressure is shown in Fig.~\ref{figbendstraight}f.
The comparison with experimental data is fairly good.\\

\noindent
In a structure constituted by an assembly of straight beams, the  bending stiffness $B_\parallel$ along the seams is very weak as it mainly involves bending of the seams, while the inflated region only rotates around these hinges. 
The bending stiffness of  the seams scales classically as $Et^3$ which is orders of magnitude smaller than the stiffness in the transverse direction ($Ew^2t$ with $w\gg t$). 
This strong stiffness anisotropy should be avoided to preserve the overall stability and stiffness of the object. 
Soft structural modes may indeed cause the failure of the structure. 
Note that this soft mode is possible because the sealed lines are rectilinear, providing the rotation axis. 
Hence, some variations are needed in the direction of the network of heat-sealed lines, such that a stiff mode will be encountered for any direction and position of bending solicitation.

\subsection{"Zigzag" Patterns}

 \begin{figure}[!ht]
  \centering
    \includegraphics[width=0.75\textwidth]{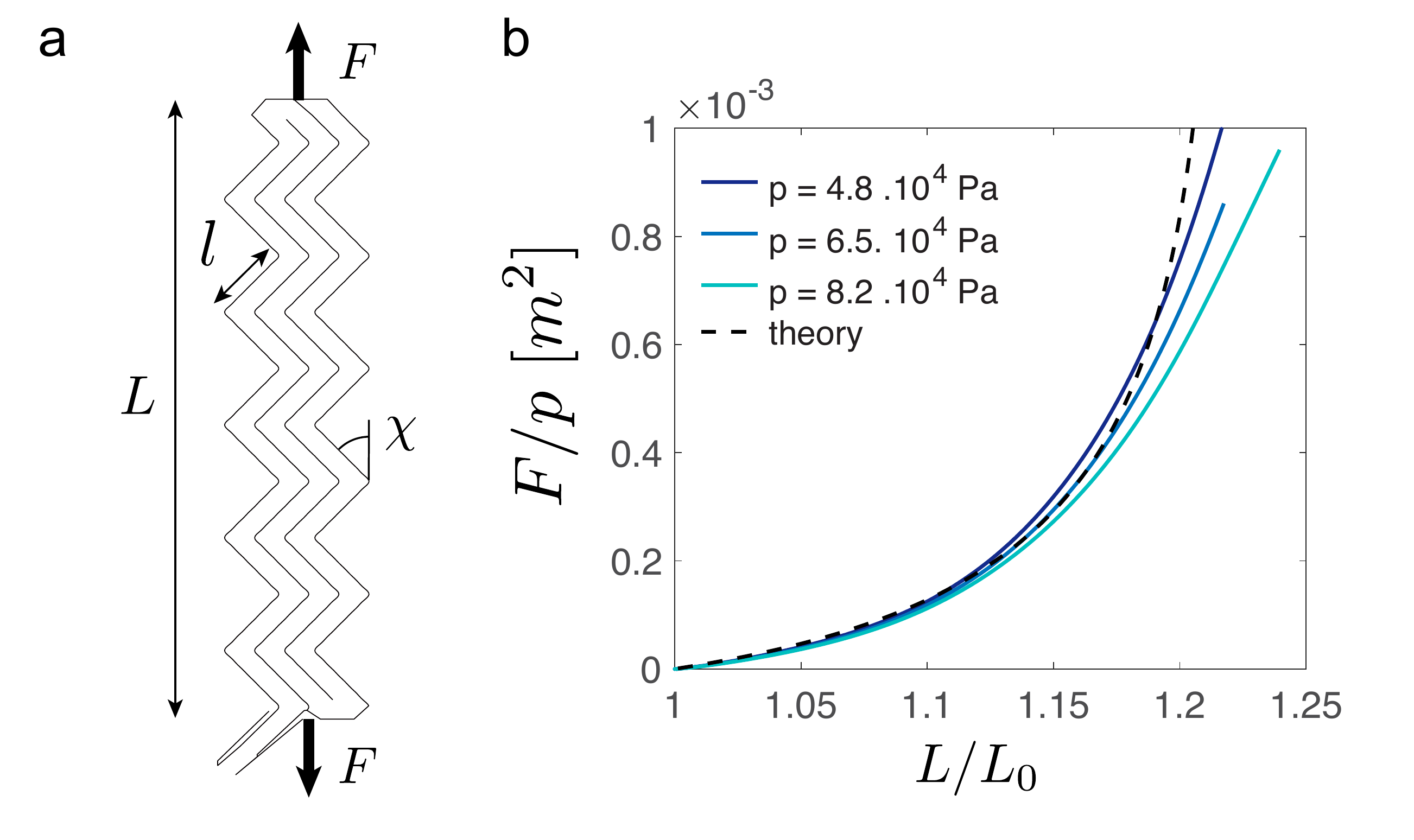}
      \caption{
      Stretching stiffness of a Zigzag pattern. (a) Sketch of the heat-sealed pattern and of the traction test. 
      (b) Theoretical (dashed line) and experimental (solid lines) force divided by the pressure as a function of the elongation. 
      The seam width of the line $e$, difficult to measure, has been used as a fitting parameter ($w=7.2\,$mm, $n=33$, $l=34$\,mm, $\chi=\pi/4$). } 
       \label{figzigzagstretch}
\end{figure}
\paragraph{Stretching.} 
As a force balance approach is not trivial in the case of the stretching of zigzag patterns (Fig. \ref{figzigzagstretch}a), we propose an energy approach to estimate the stretching rigidity.
We restrain ourselves to the ``inextensible" regime, for which the total 
{potential} energy reads $U=-pV-F\Delta$. \\
Using the same notations as in the previous section, the expression of the volume of a zigzag structure is: 
\begin{equation}
    V=\left(1-\frac{\sin{\beta}}{\beta}\right)\frac{w^2nl}{\beta}
\end{equation}
where $n$ is the number of zigzag segments and $l$ is the length of a single segment.
The contraction rate perpendicular to the local seams $\lambda$ may also be expressed as a function of the spanned angle $\beta$ (defined in Fig. \ref{figstraightstretch}):
\begin{equation}
    \lambda=2\sin{(\beta/2)}/\beta
\end{equation}
The displacement $\Delta$ along (or perpendicular to) the zigzag direction reads: 
\be
\Delta=(\lambda_\parallel-\lambda_\parallel^0)L_0
\ee
where $\lambda_\parallel=\dfrac{\lambda}{\sqrt{\sin^2{\chi}+\lambda^2\cos^2{\chi}}}$, $\lambda_\parallel^0$ is the contraction in the unstretched fully inflated state, that is, for $\beta=\pi$, and $L_0$ the length of the inflated structure in the rest state. 
\noindent
Energy minimization gives us the following expression for the force:
\begin{equation}
    F=-p\,\frac{\partial V/\partial \beta}{\partial \Delta/\partial \beta}
\end{equation}
which becomes, after derivation:
\begin{equation}
    F=2p(w-e)^2nl\cos{(\beta/2)}\,\frac{(\tan^2\chi+\sinc^2(\beta/2))^{3/2}}{\beta\tan\chi}
    \label{eq:zigstretch}
\end{equation}
The ratio of the experimental force by the applied pressure $F/p$ is plotted as a function of the elongation $L/L_0$
for various values of the pressure (Fig. \ref{figzigzagstretch}b).
All curves collapse onto one single master curve (at least for small displacements), that is very well fitted by the inextensible theory presented in equation (\ref{eq:zigstretch}), using once again the seam thickness $e$ as a single fitting parameter.\\

\noindent
Note that as in the case of miura-ori structures \cite{schenk2013geometry},  inflated zigzag patterns exhibit an auxetic behaviour: the Poisson ratio is negative around the inflated unstretched state ($\beta=\pi$) and reads:
\begin{equation}
    \nu=-\frac{\lambda^2}{\tan^2{\chi}}
\end{equation}
This expression is only valid in the inextensible regime, i.e., when $\chi$ is sufficiently large. 

\bibliographystyle{unsrt}
\bibliography{article.bib}